\documentclass[12pt]{article}
\pdfoutput=1
\usepackage{mathrsfs}
\usepackage{amsmath,amsfonts,amssymb,amsbsy,dsfont}
\usepackage{graphicx,color}
\usepackage{graphics}
\usepackage[nosort]{cite}

\newcommand{\be}{\begin{equation}}
\newcommand{\ee}{\end{equation}}
\newcommand{\bea}{\begin{eqnarray}}
\newcommand{\eea}{\end{eqnarray}}
\newcommand{\ba}{\begin{eqnarray}}
\newcommand{\ea}{\end{eqnarray}}
\newcommand{\nn}{{\nonumber}}
\newcommand{\vev}[1]{{\left< {#1} \right>}}

\newcommand{\mt}[1]{\textrm{\tiny #1}}

\newcommand{\gym}{g_\mt{YM}}

\newcommand{\email}[1]{\vbox{\center\tt#1}\vspace{3mm}}

\begin{document}
\begin{titlepage}

\rightline{\small{\tt }}
\begin{center}
\vskip 1cm
\centerline{{\Large {\bf Cusped Wilson lines in symmetric representations}}}
\vskip 1cm

{Diego H. Correa, Fidel I. Schaposnik Massolo }

{\it Instituto de F\'isica La Plata, CONICET,
Universidad Nacional de La Plata

C.C. 67, 1900 La Plata, Argentina
}
\email{correa,\;
fidel.s@fisica.unlp.edu.ar}

and
\vskip 3mm

{Diego Trancanelli}

{\it Institute of Physics, University of S\~ao Paulo

05314-970 S\~ao Paulo, Brazil
}
\email{ dtrancan@usp.br}

\vskip 3cm

{\bf Abstract}

\end{center}
\noindent
We study the cusped Wilson line operators and Bremsstrahlung functions associated to particles transforming in the rank-$k$ symmetric representation of the gauge group $U(N)$ for ${\cal N} = 4$ super Yang-Mills. We find the holographic D3-brane description for Wilson loops with internal cusps in two different limits: small cusp angle and  $k\sqrt{\lambda}\gg N$. This allows for a non-trivial check of a conjectured relation between the Bremsstrahlung function and the expectation value of the 1/2 BPS circular loop in the case of a representation other than the fundamental. Moreover, we observe that in the limit of $k\gg N$, the cusped Wilson line expectation value is simply given by the exponential of the 1-loop diagram. Using group theory arguments, this eikonal exponentiation is conjectured to take place for all Wilson loop operators in symmetric representations with large $k$, independently of the contour on which they are supported.

\end{titlepage}
\tableofcontents


\section{Introduction}

Wilson loop operators defined along a cusped line lead to the notion of cusp anomalous dimension \cite{polyakov}. This quantity, relevant in any gauge theory,  is defined as the coefficient of the logarithmic divergence that appears in the operator's expectation value because of the presence of a cusp in the contour. Physically, the cusp anomalous dimension encodes the phase acquired by a heavy particle at rest that receives a sudden kick and is set in uniform motion, and therefore contains information about the radiation emitted by such particle.

\begin{figure}
\centering
\includegraphics[width=12cm]{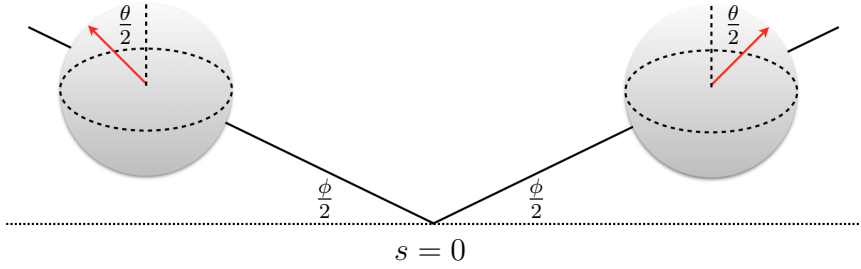}
\put(-290,81){$\tfrac{\theta}{2}$}
\put(-60,81){$\tfrac{\theta}{2}$}
\put(-215,15){$\tfrac{\phi}{2}$}
\put(-135,15){$\tfrac{\phi}{2}$}
\put(-185,-5){$s=0$}
\caption{The Wilson line operator contour (\ref{definition-WL}). The angle $\phi$ is a geometric cusp in the physical space while the angle $\theta$ is a cusp in the internal space of scalar couplings, represented by a sphere.}
\label{WLdefinition}
\end{figure}

In ${\cal N}= 4$ super Yang-Mills theory, Wilson loops are defined in terms of both the gauge field $A_\mu$ and the six scalars $\Phi_A$ of the gauge multiplet \cite{malda,rey}. It is then natural to introduce the possibility of a cusp in the internal space parametrized by the scalars. The cusp anomalous dimension can thus be defined for a geometric cusp angle $\phi$ in the contour in physical space, described by the coordinates $x^\mu$, and/or for an internal cusp angle $\theta$ in the contour in the internal space, described by the scalar couplings $n^A$. Moreover, in order to account for different types of external particles, different representations of the gauge group can be considered, so that one can define the cusp anomalous dimension (in Euclidean signature) via
\be
\frac{1}{{\rm dim} ({\cal R})} \langle {\rm tr}_{\cal R} P \exp \oint \left(i A_\mu \dot x^\mu  + |\dot x|  \Phi_A n^A \right)ds \rangle
= e^{-\Gamma_{\rm cusp}({\cal R}, \phi,\theta) \log\left(\frac{\Lambda_\mt{IR}}{\Lambda_\mt{UV}}\right)}\,.
\ee
The trace is taken in the representation ${\cal R}$ of the gauge group, which we take to be $U(N)$, and the contours in the physical and internal spaces are given by
\ba
x^\mu(s) = \left(s \cos\tfrac{\phi}{2}, -|s|\sin\tfrac{\phi}{2},0,0\right)\,,\qquad
 n^A(s) = \left(\cos\tfrac{\theta}{2}, {\rm sgn}(s)\sin\tfrac{\theta}{2},0,0,0,0\right)\,,
\label{definition-WL}
\ea
with $s$ being the parameter along the loop, see Fig.~\ref{WLdefinition}.
This ${\cal N}= 4$ super Yang-Mills cusp anomalous dimension has received considerable attention recently, but mostly limited to the case of particles in the fundamental representation. Some of the studies in this respect can be found in \cite{Drukker:2011za,CHMS,Fiol:2012sg,Correa:2012nk,Drukker:2012de,Correa:2012hh,Henn:2012qz,Gromov:2012eu,Fiol:2013iaa,Henn:2013wfa,Beccaria:2013lca,Sizov:2013joa,Bajnok:2013sya,Lewkowycz:2013laa,Fiol:2014vqa}.\footnote{In three dimensions there is a similar story for the ABJM theory, see for example \cite{Forini:2012bb,Griguolo:2012iq,Bianchi:2014laa,Correa:2014aga,Bianchi:2014ada,Aguilera-Damia:2014bqa}.}

In very specific limiting cases, exact gauge theory results can be obtained, and their strong coupling limits have been verified by explicit string theory computations. One of those limiting cases was the small cusp angles limit $(\phi,\theta \ll 1)$, which gives a near BPS operator to which supersymmetric localization \cite{Pestun} can be applied. In this limit one obtains \cite{CHMS}
\be
\Gamma_{\rm cusp}({\cal R})=(\theta^2-\phi^2) B_{\cal R}(\lambda,N) +{\cal O}(\phi^4,\theta^4)\,,
\label{Bdefinition}
\ee
where $\lambda=\gym^2N$ is the 't Hooft coupling and $B_{\cal R}(\lambda,N)$ is related to the Bremsstrahlung radiation of an accelerated charge, which can be computed in terms of the expectation value of  the 1/2 BPS circular Wilson loop
\be
B_{\cal R}(\lambda,N)=\frac{1}{2\pi^2}\lambda \,\partial_\lambda \vev{W_{\cal R}({\rm circle})}\,.
\label{relationCHMS}
\ee
In particular, for the fundamental representation an exact (in $\lambda$ and $N$) expression for the Bremsstrahlung function can be obtained \cite{CHMS,Fiol:2012sg}
\be
B_\square(\lambda,N)=\frac{\lambda}{16\pi^2 N}\frac{L_{N-1}^2\left(-\frac{\lambda}{4N}\right)+L_{N-2}^2\left(-\frac{\lambda}{4N}\right)}{L_{N-1}^1\left(-\frac{\lambda}{4N}\right)}\,,
\ee
where $L_n^\alpha(z)$ are generalized Laguerre polynomials. Another interesting regime for a fundamental charge is to consider an imaginary internal cusp such that $i\theta \gg 1$, because in this limit ladder diagrams dominate in the computation of the cusp anomalous dimension \cite{Correa:2012nk}, and they can be resummed by solving a Bethe-Salpeter equation \cite{Erickson:1999qv}.

A natural thing to do is to try to extend the study of the cusp anomalous dimension to particles that transform in representations other than the fundamental. If one considers higher rank representations, it is well understood that the dual objects to the Wilson loop are probe D-branes extending in the AdS$_5$ bulk and pinching off at the boundary along the Wilson loop contour. In particular, for Wilson loops in a totally symmetric representation of rank $k$, the string theory dual are D3-branes\footnote{Polyakov loops in higher rank representations may also be studied in terms of D-branes \cite{HK2,GKS}.} \cite{DF,GP}. The aim of this note is to study such D3-branes corresponding to straight Wilson loops with an internal cusp angle $\theta$ in the classical limit, both on the gauge theory side via perturbative expansions, and on the gravity side by constructing explicit cusped D3-brane solutions. The operators we are interested in are generically non-supersymmetric, so that one really has to solve second order equations of motion, rather than first order BPS equations. The equations of motion for the brane embeddings turn out to be highly non-linear and complicated,  and we could not solve them for generic values of the cusp angle $\theta$ and the representation rank $k$. We have however been able to solve them in two particular limits, which are interesting as they allow for a comparison with some exact gauge theory results.

The first limit we have considered is the small internal cusp angle regime, $\theta\ll1$, for which we have found a D3-brane solution valid for any value of the rank $k$.\footnote{Of course, $k$ has to be less than $N^2$ in order to have a probe brane in a fixed background.} By computing the on-shell value of the (renormalized) D3-brane action, we have been able to verify that (\ref{relationCHMS}) does indeed hold for ${\cal R}=S_k$, thus performing a test of this conjectured relation for a representation different from the fundamental one.

The second limit we have considered is the one when $k\sqrt{\lambda}\gg N$. Comparing the result obtained from the D3-brane on-shell action in this limit to a perturbative analysis at weak coupling, we have observed that the strong coupling result is simply given by the exponential of the 1-loop ladder diagram. We have checked, using group theoretical arguments, that Feynman diagrams with internal vertices are subleading in this particular limit and, moreover, that the $\ell$-th loop ladder diagram is proportional to the 1-loop ladder to the power $\ell$. In this diagrammatic derivation, we have not relied on the geometry of the loop, but only on the behavior of the Casimir coefficients of $S_k$ in the limit of $k\gg N$. This observation leads us then to the conjecture that this eikonal exponentiation might in fact be more general. We therefore propose that it should be
\be
\boxed{\vev{W_{S_k}} \simeq \exp\langle W^{({\rm 1-loop})} \rangle\,,\qquad k\gg N}\,,
\label{conjecture}
\ee
independently of the Wilson loop contour. In the discussion section we shall provide further arguments in support of this claim.


\section{D3-brane description}
\label{D3internal}

The goal of this section is to find the probe D3-brane that is the holographic dual to a cusped Wilson line in the symmetric representation $S_k$ of the group $U(N)$. We limit our analysis to the case of an internal cusp angle $\theta$, which corresponds to a non-trivial embedding of the brane on the spherical part of the metric. This case is already sufficient to perform the tests we have in mind.

\begin{figure}
\centering
\def\svgwidth{11cm}
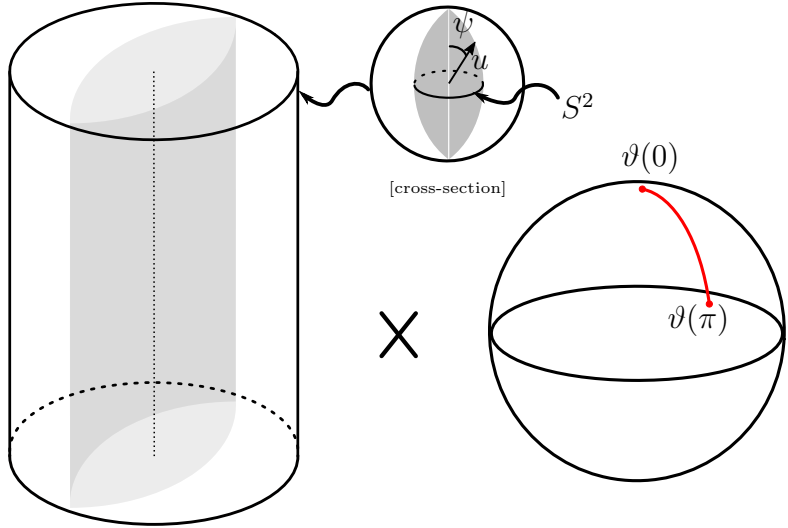
\put(-169,126){\mt{[cross-section]}}
\caption{The D3-brane with internal cusp angle. The world-volume, shaded in gray, is parametrized by the coordinates $(\tau,\psi,\chi,\xi)$. The brane reaches the AdS boundary at $\psi=0,\pi$, where the $S^2$ shrinks, as indicated in the cross-sectional view of a slice at constant  $\tau$. The internal cusp angle is given by $\theta = \vartheta(\pi) -\vartheta(0)$ on the $S^5$.}
\label{fig-brane_alt}
\end{figure}

It is convenient to use the following coordinates for the relevant components of the ${\rm AdS}_5 \times S^5$ metric:
\be
\frac{ds^2}{L^2}=\frac{du^2}{1+u^2}+(1+u^2)d\tau^2+u^2\left(d\psi^2+\sin^2\psi\, d\Omega_2^2\right)+d\vartheta^2\,.
\label{metric-internal}
\ee
Here $L$ is the AdS radius, $u$ is related to the usual global AdS radial coordinate $\rho$ by $u=\sinh\rho$, and $d\Omega_2^2=d\chi^2+\sin^2\chi \, d\xi^2$. Of the $S^5$ factor of the metric, we keep only the polar angle $\vartheta$ and suppress an $S^4$, which does not play any role here. We take the D3-brane world-volume coordinates to be $(\tau,\psi,\chi,\xi)$, with the remaining coordinates being functions of $\psi$: $u=u(\psi)$, $\vartheta=\vartheta(\psi)$. There is also a world-volume gauge field $F_{\psi\tau}=A_\tau'(\psi)$, whose associated conserved charge encodes the rank $k$ of the symmetric representation. The AdS boundary is reached at $\psi=0$ (which, upon a conformal map of the line to the cylinder, corresponds to the semi-infinite half of the operator's contour between $-\infty$ and 0) and $\psi = \pi$ (which corresponds to the semi-infinite half between 0 and $\infty$). The 2-sphere in the brane world-volume shrinks smoothly along these two lines on the boundary and the internal cusp angle is determined by the boundary value of the polar angle embedding function on the $S^5$: $\theta = \vartheta(\pi) -\vartheta(0)$. The brane is depicted in Fig.~\ref{fig-brane_alt}, where the world-volume of the brane is shaded in gray.

The DBI and WZ terms of the action (with the sign in the square root appropriate for a brane with Euclidean world-volume) read
\bea
S_\mt{DBI}&=&\frac{2N}{\pi}\int d\tau d\psi\, u^2\sin^2\psi \sqrt{(1+u^2)\left(u^2+\frac{(u')^2}{1+u^2}+(\vartheta')^2\right)+\frac{4\pi^2}{\lambda}F_{\psi\tau}^2} \,,\cr
S_\mt{WZ}&=&-\frac{2N}{\pi}\int d\tau d\psi\, u^4 \sin^2\psi\,,
\label{action-internal}
\eea
where the prime indicates derivation with respect to $\psi$. To arrive at these expressions we have used that the D3-brane tension is $T_\mt{D3}=\frac{N}{2\pi^2 L^4}$ and that, according to the AdS/CFT dictionary, $\lambda = \frac{L^4}{{\alpha'}^2}$. We have also integrated over the $S^2$.

There are two obvious conserved quantities, namely the momenta conjugated to $\vartheta$ and $A_\tau$, which are given respectively by
\bea
\Pi_\vartheta &=& \frac{1}{N} \frac{\partial{\cal L}}{\partial{\vartheta'}}= \frac{2}{\pi}\frac{u^2\sin^2\psi(1+u^2)\vartheta'}{\sqrt{(1+u^2)\left(u^2+\frac{(u')^2}{1+u^2}+(\vartheta')^2\right)+4\pi^2 F_{\psi\tau}^2/\lambda}}\,,\cr
i\Pi_A&=&  \frac{1}{N} \frac{\partial{\cal L}}{\partial{A_\tau'}} = \frac{8\pi}{\lambda}\frac{u^2\sin^2\psi F_{\psi\tau}}{\sqrt{(1+u^2)\left(u^2+\frac{(u')^2}{1+u^2}+(\vartheta')^2\right)+4\pi^2 F_{\psi\tau}^2/\lambda}}\,.
\label{conjmom-internal}
\eea
Here we have defined the second momentum with an $i$ to get a real quantity (in the Euclidean theory, the gauge field is in fact imaginary). The definitions above can be inverted and the solutions for $\vartheta'$ and $F_{\psi\tau}$ in terms of the two constants can be replaced into the equations of motion. As a result, we obtain a single but highly non-linear and complicated differential equation for $u(\psi)$:
\bea
0&\!=\!&
-16u^2 \sin^2\psi \left(u^2+1\right) \left(u'^2 +u^4+u^2\right)^2 \nn\\
&& \quad \times
\sqrt{\frac{\lambda  {\Pi_A}^2+ 16 u^6
       \sin^4\psi + 16 u^4 \sin^4\psi- 4\pi^2 \Pi_\vartheta^2 +
       \lambda  {\Pi_A}^2 u^2}{\left(u^2+1\right) \left(u'^2+u^4+u^2\right)}}
   \cr  &&
  + u^6 \left(5\lambda{\Pi_A}^2+160 \sin^4\psi\, u'^2+ 48 \sin^4\psi-4\pi^2 \Pi_\vartheta^2\right)
   \cr   &&
   +4u^4\left(\lambda{\Pi_A}^2+u'^2\left(\lambda {\Pi_A}^2+16\sin^4\psi\right)-2\pi^2 \Pi_\vartheta^2\right)
   \cr   && +2 u^8\left(\lambda {\Pi_A}^2+48\sin^4\psi\, u'^2+80\sin^4\psi \right)
\cr   &&
   -2 u^3\left(16\sin^3 \psi \cos\psi\, u'^3+\left(\lambda {\Pi_A}^2-2 \pi ^2 \Pi_\vartheta^2\right) u''\right)
   \cr   &&
   -u^5
   \left(u'' \left(\lambda \Pi_A^2+16\sin^4\psi\right)+32\sin^3\psi\cos\psi\, u'^3+32\sin^3\psi \cos \psi\, u'\right)
   \cr &&
   -16 u^9 \sin^3\psi \left(\sin\psi\, u''+2\cos\psi\, u'\right)-32u^7\sin^3\psi \left(\sin\psi\, u''+2 \cos\psi\, u'\right)
   \cr   &&
   +64 u^{12} \sin^4\psi +176 u^{10} \sin^4\psi +u\left(4 \pi^2 \Pi_\vartheta^2-\lambda{\Pi_A}^2\right) u''
   \cr  &&
  +u^2 \left(\lambda \Pi_A^2- 4\pi^2 \Pi_\vartheta^2 + 6 \left(\lambda {\Pi_A}^2-2 \pi^2 \Pi_\vartheta ^2\right) u'^2\right)
  +2 \left(\lambda {\Pi_A}^2-4 \pi ^2 \Pi_\vartheta ^2\right) u'^2\,.
\eea
In the following, we will solve this differential equation in two different limits.


\subsection{Small cusp angle limit}

In this subsection we consider the small internal cusp angle limit, $\theta\ll 1$. As we will see below, this corresponds to small $\Pi_\vartheta$. For $\theta=0$ (corresponding to the brane always sitting at a point of the $S^5$), we should recover the 1/2 BPS D3-brane solution of \cite{DF}, which is given in terms of the parameter
\be
\kappa=\frac{k\sqrt{\lambda}}{4N}\,,
\label{kappa}
\ee
where $k$ is an integer giving the `string charge' on the brane world-volume, and is identified with the rank of the symmetric representation on the gauge theory side. We make the following ansatz for the embedding
\be
u(\psi)=\frac{\kappa}{\sin\psi}\left(1+ \Pi_\vartheta F_1(\psi)+ \Pi_\vartheta^2 F_2(\psi)
+ \mathcal{O}(\Pi^3_\vartheta)\right)\,,
\label{embedding-internal}
\ee
which is basically an expansion around the 1/2 BPS limit, given by $u(\psi)=\kappa/\sin\psi$ \cite{DF}. This limit also fixes
\be
\Pi_A=\frac{4\kappa}{\sqrt{\lambda}}=\frac{k}{N}\,,
\label{PiA}
\ee
which is what we use in the following. Expanding the equation of motion of $u(\psi)$ for small $\Pi_\vartheta$, one finds
\ba
2 \sin2\psi F_1'(\psi) + (1 + 2 \kappa^2 - \cos2\psi) F_1''(\psi) &=& 0\,,
\nn\\
2 \sin2\psi F_2'(\psi) + (1 + 2 \kappa^2 - \cos2\psi) F_2''(\psi) &=& \frac{\pi^2}{1 + 2 \kappa^2 - \cos2\psi}\,,
\ea
which are solved by
\be
F_1(\psi)=0\,,\qquad F_2(\psi)=\frac{\pi^2}{8\kappa^2(1+\kappa^2)}\left(\arctan^2\left(\frac{\kappa\cot\psi}{\sqrt{1+\kappa^2}}\right)-\frac{\pi^2}{4}\right)\,.
\label{smallcuspsolution}
\ee
The constants of integration in these functions have been fixed such that
\be
F_i(0)=0\,,\qquad F'_i(\pi/2)=0 \qquad\text{for}\qquad i=1,2\,.
\ee
The first condition enforces that, at the boundary, the brane profile reproduces the 1/2~BPS brane of \cite{DF}, whereas the second condition guarantees a smooth brane profile, with no cusps at the midpoint. Notice, however, that the function $F_2(\psi)$ in (\ref{smallcuspsolution}) can be used to represent the D-brane configuration in the full range $0\le \psi\le \pi$.

At this point we want to evaluate the action (\ref{action-internal}) on-shell, after expanding it for small $\Pi_\vartheta$. One finds that
\be
S_\mt{DBI}+S_\mt{WZ}=-TN\frac{\pi^2}{4}\frac{\kappa}{(1+\kappa^2)^{3/2}} \Pi_\vartheta^2 + \mathcal{O}(\Pi_\vartheta^4)\,,
\ee
where $T=\int d\tau$ is an infrared cutoff on the $\tau$-extension of the brane. This result is finite, but we still need to include appropriate boundary terms for the fields. These boundary terms have the form of a Legendre transform
\be
S_{\rm bdry}=
-\int d\tau \left(u \,\Pi_u +A_\tau \,\Pi_A\right)\Big|^{\psi=\pi}_{\psi=0}\,, 
\ee
and alter the boundary conditions of the fields exchanging Neumann and Dirichlet boundary conditions, see \cite{DGO,DF}. The result is
\be
S_{\rm bdry} = T N\frac{\pi^2}{4}\frac{1+2\kappa^2}{\kappa(1+\kappa^2)^{3/2}} \Pi_\vartheta^2 + \mathcal{O}(\Pi_\vartheta^4)\,,
\ee
so collecting the two contributions we get
\be
\log\vev{W_{S_k}} = -TN \frac{\pi^2}{4} \frac{1 }{\kappa\sqrt{1+\kappa^2}}  \Pi_\vartheta^2+ \mathcal{O}(\Pi_\vartheta^4)\,.
\ee
Now we need to convert $\Pi_\vartheta$ into the internal cusp angle, via
\be
\theta=\int^\pi_0d\psi\, \vartheta'(\psi)\,.
\ee
This relation, to leading order in the small cusp angle limit, turns out to be
\be
\Pi_\vartheta=\frac{2\kappa\sqrt{1+\kappa^2}}{\pi^2}\theta+ \mathcal{O}(\theta^2)\,,
\ee
which finally leads to
\be
\log\vev{W_{S_k}} = -T\frac{N}{\pi^2}\kappa\sqrt{1+\kappa^2}\, \theta^2 + \mathcal{O}(\theta^4)\,.
\ee
This means that the strong coupling limit of the Bremsstrahlung function is
\be
B_{S_k}\simeq\frac{N}{\pi^2}\kappa\sqrt{1+\kappa^2} \,,\qquad [\,
N,\lambda\gg 1\,,\quad \text{any}\; \kappa \, ]\,.
\label{BSkstrong}
\ee
The same strong coupling limit of the Bremsstrahlung function for charges in a totally symmetric representation can be read off from previous computations in \cite{Fiol:2011zg}, where a charge following a hyperbolic trajectory was considered. More recently, D3-branes dual to charges following arbitrary time-like trajectories were constructed in \cite{Fiol:2014vqa} and the computation of their energy loss is also consistent with our result.


\subsection{Large $\kappa$ limit}
\label{largekappasugra}

A second interesting limit to consider is the one of $\kappa\gg 1$. If we consider the large $\kappa$ limit of the small cusp angle solution found in the previous subsection we find
\be
u(\psi)=\frac{\kappa}{\sin\psi}\left(1 - \frac{\pi^2}{8}\psi(\pi-\psi)\frac{\Pi_\vartheta^2}{\kappa^4}
+\ldots\right) \,,
\label{largekappasmalltheta}
\ee
and see that the corrections are organized in powers of $\frac{\Pi_\vartheta}{\kappa^2}$. This inspires the definition of a new expansion parameter
\be
\pi_\vartheta = \frac{\pi^2}{2\kappa^2}\Pi_\vartheta\,,
\ee
which will be kept fixed as $\kappa$ and $\Pi_\vartheta$  go to infinity. We consider now arbitrary angles $\theta$ and make the ansatz
\be
u(\psi)=\frac{\kappa}{\sin\psi} F(\psi)\,,
\ee
which leads to the following equation of motion, valid for large $\kappa$ and fixed $\pi_\vartheta$,
\be
4F^3 F'' - \pi^2 \pi_\vartheta^2 = 0\,.
\ee
Its solution with $F(0)=F(\pi) =1$ and such that $F(\psi) \to 1$ as $\pi_\theta\to 0$ is
\be
F(\psi) = \frac{1}{\pi} \sqrt{\pi^2-2\psi(\pi-\psi)\left(1-\sqrt{1-\pi_\vartheta^2}\right)}\,.
\ee
We can now relate  $\pi_\vartheta$ with the cusp angle $\theta$. For $0\le\pi_\vartheta\le 1$
\bea
\theta &=& \int_{0}^{\pi} d\psi\, \vartheta'(\psi) = \int_{0}^{\pi} d\psi \frac{\pi\ \pi_\vartheta}{\pi^2 - 2\psi(\pi-\psi)\left(1-\sqrt{1-\pi_\vartheta^2}\right)} +{\cal O}(\kappa^{-1})\cr
&=& \arcsin \pi_\vartheta  +{\cal O}(\kappa^{-1})\,.
\eea

As before, we want to compute the action (\ref{action-internal}) on-shell, which in the large $\kappa$ limit reduces to
\bea
S_\mt{DBI}+S_\mt{WZ} &=& -T \frac{2N\kappa^2}{\pi^2} \left(1-\sqrt{1-\pi_\vartheta^2}\right)
+{\cal O}(\kappa)\cr
&=& -T \frac{2N\kappa^2}{\pi^2} (1-\cos\theta) +{\cal O}(\kappa)\,.
\eea
For the boundary terms we have
\be
S_{\rm bdry} = T  \frac{4N\kappa^2}{\pi^2} (1-\cos\theta) +{\cal O}(\kappa)\,,
\ee
and finally
\be
\log\vev{W_{S_k}} = -T  \frac{2N\kappa^2}{\pi^2} (1-\cos\theta) +{\cal O}(\kappa)\,.
\ee
The cusp anomalous dimension for generic angle $\theta$ and large $\kappa$ is therefore given by
\be
\Gamma_{\rm cusp} (S_k)= \frac{2N\kappa^2}{\pi^2} (1-\cos\theta)\,,\qquad [\, \kappa\gg 1\,,\quad N,\lambda\gg1\,,\quad {\rm any } \;\theta\,]\,.
\label{gammastrong}
\ee

From (\ref{gammastrong}) we can extract the Bremsstrahlung function, by expanding for small cusp angle $\theta$. The result coincides, of course, with the large $\kappa$ limit of (\ref{BSkstrong}) and gives
\be
B_{S_k} = \frac{k^2\lambda}{16\pi^2 N}\,,\qquad [\,N,\lambda\gg 1\,]\,.
\ee
This is in perfect agreement with the gauge theory prediction obtained in \cite{FT} from supersymmetric localization. This can be seen as a non-trivial test that the conjecture (\ref{relationCHMS}) of \cite{CHMS} is valid for representations other than the fundamental one.


\section{Perturbative cusp anomalous dimension}
\label{perturbative}

We now want to connect the results obtained in the previous section from supergravity with a perturbative analysis at small 't Hooft coupling. In particular, we are going to focus on Wilson lines in the symmetric representation and with large $\kappa$. At small coupling, large $\kappa$ implies that $k\gg N$. We are going to find that in this limit the ladder diagrams are the only ones that contribute to the operator's expectation value. Moreover, the full perturbative series exponentiates, so that the cusp anomalous dimension in (\ref{gammastrong}) turns out to be simply given by the leading order result at weak coupling. This exponentiation is seemingly only due to the behavior of the color factors in this limit, and does not depend on the specific geometry of the loop, an observation which has led us to propose the conjecture in (\ref{conjecture}).

The analysis starts by expanding the exponential in the Wilson line in (\ref{definition-WL}) to quadratic order in the fields. This gives the vacuum expectation value of the operator to 1-loop order\footnote{Note that the factor of $1/2!$ coming from the Taylor expansion of the exponential is compensated by picking a particular path ordering, in this case $s_2\ge s_1$. This is true at all orders in the expansion.}
\be
\langle W_{\cal R} \rangle \simeq  1 + \frac{{\rm tr}(T^aT^b)}{\dim{({\cal R})}}\! \int_{-\infty}^\infty \!\! ds_1\!
\int_{s_1}^{\infty} \!\!ds_2\left( n^A_1 n^B_2 \langle \Phi_A^a(x_1) \Phi_B^b(x_2)\rangle-
\dot x^\mu_1 \dot x^\nu_2 \langle A^a_\mu(x_1) A^b_\nu(x_2)\rangle\right)\,.
\ee
For compactness, we have used the shorthands $x_i = x(s_i)$ and $n_i = n(s_i)$. The generators $T^a$ are in the representation ${\cal R}$ of the group $U(N)$. Using the propagators in Feynman gauge
\be
\langle \Phi^a_A(x) \Phi^b_B (y) \rangle = \frac{g_\mt{YM}^2}{4\pi^2}\frac{\delta^{ab}\delta_{AB}}{|x-y|^2}\,,
\qquad
\langle A^a_\mu (x) A^b_\nu (y) \rangle = \frac{g_\mt{YM}^2}{4\pi^2}\frac{\delta^{ab} \delta_{\mu\nu}}{|x-y|^2}\,,
\ee
it is easy to see that the net contribution comes only from $s_1\leq 0$ and $s_2\geq 0$, thus giving
\be
\langle W_{\cal R} \rangle  \simeq  1 + \frac{{\rm tr}(T^a T^a)}{\dim{({\cal R})}}\frac{g_\mt{YM}^2}{4\pi^2}
\int^{\infty}_0 ds_1\int_{0}^{\infty} ds_2 \frac{\cos\phi-\cos\theta}{s_1^2+s_2^2+2s_1 s_2\cos\phi}\,.
\label{pert-1loop}
\ee
The color factor of this 1-loop contribution is given by
\be
{\rm tr}(T^a T^a) = \dim{({\cal R})}\  \frac{C_2({\cal R})}{2}\,,
\ee
where $C_2({\cal R})/2$ is the coefficient in the quadratic Casimir operator of the representation ${\cal R}$. We explain our group theoretical conventions in more detail in the appendix. After performing the integrations, which are the same as the ones encountered for the fundamental representation \cite{ESZ}, one arrives at
\be
\langle W_{\cal R} \rangle  =  1 - \frac{\lambda}{8\pi^2}  \frac{C_2({\cal R})}{N}
 (\cos\phi-\cos\theta)\frac{\phi}{\sin\phi}\log\left(\frac{\Lambda_\mt{IR}}{\Lambda_\mt{UV}}\right)+{\cal O}(\lambda^2)\,.
\ee
The presence of a logarithmic divergence (regulated here by the introduction of cutoffs) is typical of loops which either have cusps or self-intersect \cite{DGO}.

For a totally symmetric representation of rank $k$, the 1-loop cusp anomalous dimension is then given by
\be
\Gamma_{\rm cusp}(S_k) = \frac{\lambda k(N+k-1)}{8\pi^2 N } (\cos\phi-\cos\theta)\frac{\phi}{\sin\phi} +{\cal O}(\lambda^2)\,,
\label{gammaweak}
\ee
where we have used that $C_2(S_k)=k(N+k-1)$, see the appendix. This result is valid for any values of $k$ and $N$ and for any cusp angles $\phi$ and $\theta$.

It is now interesting to consider the limit of $\kappa\gg 1$ of the result above for the symmetric representation, because in this limit we can compare with the supergravity analysis of Sec. \ref{largekappasugra}. At weak coupling, large $\kappa$ implies $k\gg N$,\footnote{We also implicitly consider $N\gg 1$ throughout this section, to be able to compare with the supergravity analysis of Sec. \ref{D3internal}.} so that setting $\phi=0$ the cusp anomalous dimension (\ref{gammaweak}) becomes
\be
\Gamma_{\rm cusp}(S_k) = \frac{\lambda k^2}{8\pi^2 N } (1-\cos\theta) + {\cal O}\left(\lambda^2,\frac{1}{k},\frac{N}{k}\right)\,.
\label{gammaweak2}
\ee
Remarkably, as we have anticipated above, this 1-loop result already coincides with the strong coupling result (\ref{gammastrong}). In what follows, we shall argue that $\Gamma_{\rm cusp}(S_k)$ receives no further corrections in the regime  $k\gg N$.

To analyze higher loop contributions it is convenient to classify them into ladder diagrams and interaction diagrams, corresponding to the cases without and with internal vertices, respectively.


\subsection{Ladder diagrams}

The 1-loop diagram computed in (\ref{pert-1loop}) is the simplest ladder diagram. For $\phi = 0$ and generic internal cusp angle $\theta$, it becomes
\be
\langle W_{\rm ladders}^{({\rm 1-loop})} \rangle =
\frac{C_2(S_k)}{2}\frac{\lambda}{4\pi^2N}
  \!\int\limits^{\infty}_0 dt_1 \int\limits_{0}^{\infty}  ds_1
  \frac{1-\cos\theta}{(s_1 + t_1)^2}\,.
\ee

At 2-loop order there are two kinds of diagrams:\footnote{Recall that whenever a propagator begins and ends in the same half-line, the corresponding contribution vanishes \cite{ESZ}.}
\ba
\label{2loopladderseq}
\langle W_{\rm ladders}^{({\rm 2-loop})} \rangle  & \!=\! &
\frac{{\rm tr}(T^a T^a T^b T^b)}{\dim{(S_k)}}\left(\frac{\lambda}{4\pi^2N}\right)^2
  \!\int\limits^{\infty}_0 dt_1 \int\limits_{0}^{\infty}  ds_1 \int\limits^{t_1}_0  dt_2 \int\limits_{0}^{s_1}  ds_2\frac{(1-\cos\theta)^2}{(s_1 + t_1)^2(s_2 + t_2)^2}
\cr
&& \! +\frac{{\rm tr}(T^a T^b T^a T^b)}{\dim{(S_k)}}\left(\frac{\lambda}{4\pi^2N}\right)^2
  \!\int\limits^{\infty}_0 dt_1 \int\limits_{0}^{\infty}  ds_1 \int\limits^{t_1}_0  dt_2 \int\limits_{0}^{s_1}  ds_2
\frac{(1-\cos\theta)^2}{(s_1 + t_2)^2(s_2 + t_1)^2}\,.\cr &&
\ea
\begin{figure}[h]
\centering
\def\svgwidth{13cm}
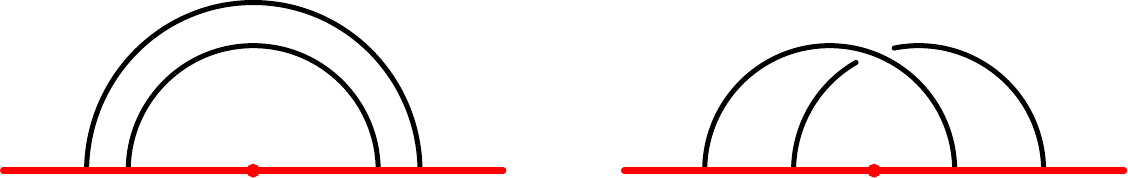
\caption{2-loop parallel (left) and crossed (right) ladder diagrams.}
\label{2loopladders}
\end{figure}
The first line in (\ref{2loopladderseq}) comes from a ladder with parallel rungs, while the second line comes from a ladder with crossing rungs, as depicted in Fig. \ref{2loopladders}. They possess different color factors produced by traces of four generators in different orders. In the case of totally symmetric representations, one can show (see appendix) that
\be
\frac{{\rm tr}(T^a T^b T^a T^b)}{{\rm tr}(T^a T^a T^b T^b)} =
1-\frac{N}{k(N+k-1)}+ \frac{1}{(N+k-1)^2}\,.
\label{color-factors}
\ee
From this formula, we see that when $k=1$, {\it i.e.} for the fundamental representation, the crossing ladder diagrams are subleading in the planar limit. However, for $k>1$, crossing ladders contribute to the same order as the parallel ladders and have to be taken into account. Doing this for the general case would be quite complicated. It is only the parallel ladder diagrams that can be resummed by solving an integral Bethe-Salpeter equation, which incidentally explains why the fundamental representation is simpler.

In the limit when $k\gg N$, a crucial simplification takes place. Up to terms of order $N/k^2$, one has in fact that
\be
{\rm tr}(T^a T^b T^a T^b) \simeq {\rm tr}(T^a T^a T^b T^b)
= \dim{(S_k)}\left(\frac{C_2(S_k)}{2}\right)^2\,,
\ee
as we show in the appendix. The fact that the two terms in (\ref{2loopladderseq}) become normalized with the same color factor makes the combined integrand symmetric in the exchange of the 1 and 2 labels. All the integrals can then be extended from minus to plus infinity and everything gets rearranged into a single term,
\ba
\langle W_{\rm ladders}^{({\rm 2-loop})} \rangle  & \!\simeq\! &
\frac{1}{2}\left(\frac{C_2(S_k)}{2}\frac{\lambda}{4\pi^2N}\right)^2
\left(\int\limits^{\infty}_0 dt_1 \int\limits_{0}^{\infty}  ds_1 \frac{1-\cos\theta}{(s_1 + t_1)^2}\right)^2\nn
\\
& \!=\! & \frac{1}{2} \langle W_{\rm ladders}^{({\rm 1-loop})} \rangle^2\,.
\label{2loopladderfinal}
\ea

At higher loop orders, more types of ladder diagrams appear. All of them come with different color factors originating from traces of generators in different orders. It is however not difficult to verify that to any loop order $\ell$ (see appendix)
\ba
{\rm tr}(T^{a_1} \cdots T^{a_\ell} T^{\sigma(a_1)} \cdots T^{\sigma(a_\ell)})
\simeq  {\rm tr}(T^{a_1}T^{a_1} \cdots T^{a_\ell} T^{a_\ell})
=
\dim{(S_k)}\left(\frac{C_2(S_k)}{2}\right)^\ell\,,
\ea
for any permutation $\sigma$ and for $k\gg N$. The leading order integrals for large $k$ of the $\ell$-loop ladder contributions can then be gathered as follow
\be
\sum_{\sigma}\prod_{i=1}^\ell \int\limits^{t_{i-1}}_0 \!dt_i \int\limits^{s_{i-1}}_0 \!ds_i \frac{1}{(s_i + t_{\sigma(i)})^2}
=\frac{1}{\ell!} \left(\int\limits^{\infty}_0 dt_1 \int\limits_{0}^{\infty}  ds_1 \frac{1}{(s_1 + t_1)^2}\right)^\ell\,,
\ee
where the sum is over all possible permutations of $\ell$ elements and $s_0=t_0=\infty$. Therefore,
\be
\langle W_{\rm ladders}^{({\rm \ell-loop})} \rangle  \simeq
\frac{1}{\ell !} \langle W_{\rm ladders}^{({\rm 1-loop})} \rangle^\ell\,,
\ee
which leads to an eikonal exponentiation of the ladder contributions at leading order in  the large $k$ limit,
\be
\langle W_{\rm ladders} \rangle=\sum_{\ell=0}^\infty \langle W_{\rm ladders}^{({\rm \ell-loop})} \rangle
\simeq \exp {\langle W_{\rm ladders}^{({\rm 1-loop})} \rangle }\,.
\ee
We stress that this result is valid for the totally symmetric representation $S_k$ with $k\gg N$.


\subsection{Interaction diagrams}

\begin{figure}[h]
\centering
\def\svgwidth{10cm}
\includegraphics[width=13cm]{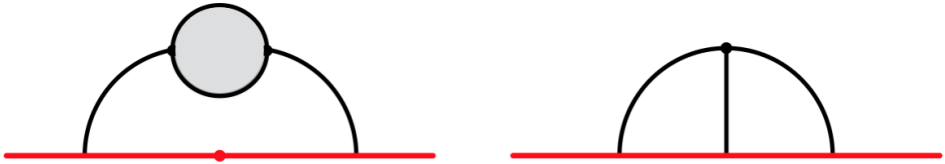}
\caption{Self-energy correction to the propagator at 2-loop order (left) and internal 3-vertex graph (right).
}
\label{2loopinteraction}
\end{figure}

To prove (\ref{conjecture}), we need to show that ladders are the only contribution to the Wilson line's expectation value. In principle, there could also be interaction diagrams, which start contributing at 2-loop order, as shown in Fig. \ref{2loopinteraction}. We are not going to compute the corresponding integrals explicitly but, by considering their associated color factors, we shall argue that all these contributions are subleading in the regime $k\gg N$.

For the 2-loop interaction diagrams we have the self-energy correction to the propagator and the diagram with one internal 3-vertex.
The self-energy correction receives contributions from all the fields of the theory plus the ghosts. We just notice that the overall color factor of this diagram becomes
\be
\frac{\gym^4}{\dim({\cal R})}{\rm tr}(R_j^i R_l^k) f^{jnq}_{imp} f^{lmp}_{knq} \propto
\left(\frac{\lambda}{N}\right)^2\left(NC_2({\cal R})-C_1({\cal R})^2\right)\,,
\label{2loopint1}
\ee
where we have used the canonical basis described in the appendix and that in this basis the structure constants are $f^{ikm}_{jln}\propto \delta^i_l\delta^k_n\delta^m_j-\delta^k_j\delta^i_n\delta^m_l$.

The internal 3-vertex diagram comes from contracting the third order term in the Taylor expansion of the exponential of the Wilson line with the two 3-vertices in the ${\cal N}=4$ super Yang-Mills action, schematically given by $f^{mnp}\left( \partial A^m A^n A^p+\partial \Phi^m A^n\Phi^p\right)$. This contraction gives the same color factor as (\ref{2loopint1})
\be
\frac{\gym^4}{\dim({\cal R})}{\rm tr}(R^j_i R^l_k R^n_m) f^{ikm}_{jln}\propto
\left(\frac{\lambda}{N}\right)^2\left(NC_2({\cal R})-C_1({\cal R})^2\right)\,.
\label{2loopint}
\ee

For totally symmetric representations with $k\gg N \gg 1$, the Casimir coefficients are $C_p(S_k) \simeq k^p $ to leading order, as reviewed in the appendix. These 2-loop interaction diagrams are therefore suppressed by a factor of $N/k^2$, if compared to the 2-ladder contribution in  (\ref{2loopladderfinal}).

\begin{figure}[h]
\centering
\def\svgwidth{15cm}
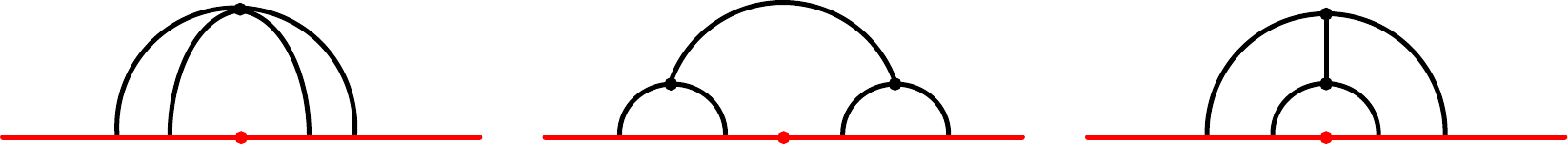
\caption{Some examples of 3-loop diagrams with four legs attached to the Wilson line.
}
\label{3loopinteraction}
\end{figure}

At higher orders, many diagrams contribute. Some representative 3-loop order diagrams with four legs on the Wilson line are depicted in Fig.~\ref{3loopinteraction}.

Generically, an $\ell$-loop diagram scales like $\gym^{2\ell}=\left(\lambda/N\right)^\ell$, a factor which is obtained from joining together $P$ propagators and $V$ vertices, with $\ell=P-V$. Out of the $P$ propagators, $P_{({\rm i})}$ will be internal, $P_{({\rm e})}$ of the external ones will connect the loop with a vertex and the remaining $P_{({\rm r})}$ will be rungs. Of the $V$ vertices, $V_{(3)}$ will be 3-vertices, each contributing a structure constant $f^{bcd}$, and $V_{(4)}$ will be 4-vertices, each contributing a factor of $f^{lmn}f^{lpq}$. The number of vertices and propagators are related to
the loop order $\ell$ by
\be
\ell=P_{({\rm i})}+P_{({\rm e})}+P_{({\rm r})}-V_{(3)}-V_{(4)}\,,\qquad 3V_{(3)}+4 V_{(4)}= 2P_{({\rm i})}+P_{({\rm e})}\,.
\label{constraintFeynman}
\ee
The color factor of these diagrams is then given by
\be
\frac{\gym^{2\ell}}{\dim({\cal R})}{\rm tr}\left(T^{a_1}\cdots T^{a_{G}}\right)
\prod_{i=1}^{V_{(3)}}f^{b_ic_id_i}\prod_{j=1}^{V_{(4)}}f^{lm_jn_j}f^{lp_jq_j}
\sum_{(r,s)\in S}\prod_{\alpha=1}^P \delta^{r_\alpha s_\alpha}\,,
\ee
where $G = P_{({\rm e})} + 2 P_{({\rm r})}$ is the number of generators and $S$ is the set of indices that can be connected by the propagators, schematically given by $S=\{(a,bcd),(a,mnpq),(bcd,mnpq)\}$, not allowing for tadpoles produced by $(r,s)=(bcd,bcd)$ and $(r,s)=(mnpq,mnpq)$. Combining the formulas in (\ref{constraintFeynman}), we see that
\be
G = 2\ell - V_{(3)} - 2V_{(4)}.
\label{G}
\ee
In general, the trace of $G$ generators will result in a sum of terms of the form
\be
\prod_{s=1}^{G} C_s({\cal R})^{q_s}\,,\qquad \forall q_s: \quad 0\le q_s\le G\,,\qquad \sum_{s=1}^G s \,q_s\le G
\,.
\ee
Then for totally symmetric representations with $k\gg N$,
\be
\prod_{s=1}^{G} C_s({\cal R})^{q_s} \lesssim k^G\,.
\label{boundapp}
\ee
From (\ref{G}) and (\ref{boundapp}) it is clear why ladder diagrams dominate at every fixed loop order $\ell$.\footnote{The bound in (\ref{boundapp}) might not be saturated when consecutive generators contracted with structure constants give rise to commutators, which can produce relative factors of $N/k$. Additional $N$ factors appear for each internal loop, but this requires additional vertices. When compared with the ladder diagram of the same loop order, the relative factor is always at least $N/k^2$.}

We can check these general statements for the specific examples of the diagrams in Fig.~\ref{3loopinteraction}. Those diagrams come with the following color factor
\be
\frac{\gym^6}{\dim({\cal R})}{\rm tr}(R^j_i R^l_k R^n_m R^q_p) f^{sik}_{tjl} f^{tmp}_{snq} \propto
\left(\frac{\lambda}{N}\right)^3 N \left(NC_2({\cal R})-C_1({\cal R})^2\right)\,,
\label{3loopinta}
\ee
which for totally symmetric representations with $k\gg N$ are suppressed by a factor $(N/k^2)^2$
when compared with the corresponding ladder diagrams.

At this order in $\lambda$, another possibility is to add a rung to the  diagrams shown in Fig.~\ref{2loopinteraction}.
\begin{figure}[h]
\centering
\includegraphics[width=11cm]{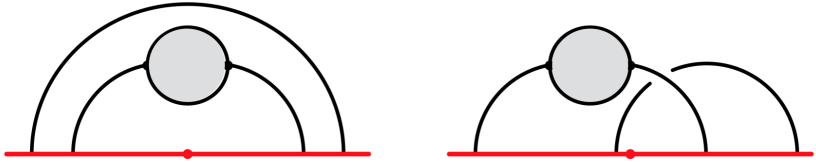}
\caption{3-loop diagrams obtained from 2-loop diagrams by adding an extra rung.
}
\label{3loopinteraction2}
\end{figure}
If the rung is parallel to the existing one (see Fig.~\ref{3loopinteraction2} - left), the factor (\ref{2loopint}) acquires an extra $\tfrac{\lambda}{N}C_2$. If the rung crosses it (see Fig.~\ref{3loopinteraction2} - right), the total factor becomes proportional to $(\tfrac{\lambda}{N})^3 (N C_2^2 + N C_1^2-N^2C_2-C_2 C_1^2)$. In any case, for $k\gg N$, both contributions are subleading by a factor $N/k^2$ if compared to the corresponding ladder diagrams at order $\lambda^3$.

Given that interaction terms do not contribute at leading order for $k\gg N$, we see that in this limit
\be
\vev{W_{S_k}}\simeq \vev{W_{\rm ladders}} \simeq \exp {\langle { W^{({\rm 1-loop})}}\rangle }\,,
\ee
and the 1-loop ladder result completely captures the expectation value of the operator!


\section{Discussion}

A few comments about our results are now in order. First of all, we should specify more precisely what we mean, in Sec.~\ref{perturbative}, by $k\gg N$. There is in fact a caveat, due to a possible back-reaction and deformation of the background ${\rm AdS}_5\times S^5$ geometry, which takes place when the probe gets `too heavy'. In the gauge theory language, this happens when the operator insertion scales with the same power of $N$ as the action, namely as $N^2$. This back-reaction is therefore avoided by taking $k$ of order ${\cal O}(N^{\nu})$ with $1<\nu<2$, which is what we have implicitly assumed.

In order to argue for the exponentiation of the 1-loop ladder result in Sec.~\ref{perturbative} we have not used any details of the geometry of the Wilson loop contour, but only general group theoretical arguments. In fact, this exponentiation is not only limited to cusped lines. In \cite{HK}, it had already been observed that it also takes place for the 1/2 BPS circle.
Notice that the circle is just a representative of an infinite family of operators (generically 1/8 BPS) that can be defined for arbitrary contours on a 2-sphere and whose expectation value is captured by the same Gaussian matrix model as the circle \cite{DGRT,pestun2}. The only difference is that the coupling constant of the matrix model is rescaled by a quantity that only depends on the geometry of the loop, and not on the representation of the operator.\footnote{More specifically, if ${\cal A}_\mt{in}$ and ${\cal A}_\mt{out}$ are the areas of the surfaces on the $S^2$ inside and outside of the loop, respectively, then the new coupling constant is given by $\gym^2\to\gym^2 {\cal A}_\mt{in}{\cal A}_\mt{out}/4\pi^2$.} The same exponentiation will therefore take place for those operators as well, which represents a non-trivial argument in favor of our claim~(\ref{conjecture}).

Interestingly, this exponentiation appears not to be limited to the symmetric representation $S_k$. Representations of $U(N)$ characterized by rectangular Young tableaux with $k$ columns and $f$ rows will also lead to the same conclusions, if we take $k\gg N$ and $f\ll N$. The Casimir operators for those representations are in fact given by \cite{barut}
\be
C_p(\underbrace{k,\ldots,k}_{f\; {\rm times}},0,\ldots,0)=f k(k+N-f)^{p-1}\simeq f k^p\,,
\ee
and scale with $k$ as in the $S_k$ case. The corresponding cusp anomalous dimensions are therefore going to be given by
\be
\Gamma_{\rm cusp}(\underbrace{k,\ldots,k}_{f\; {\rm times}},0,\ldots,0) = f\frac{\lambda k^2}{8\pi^2 N } (\cos\phi-\cos\theta)\frac{\phi}{\sin\phi}\,,\qquad [\,k\gg N\,,\quad f\ll N\,]\,,
\label{generalization}
\ee
which is a non-trivial gauge theory prediction for a non-Abelian DBI action computation on the string theory side. In fact, the requirement itself that the gauge group be $U(N)$ might also be relaxed. Rank-$k$ totally symmetric representations of $O(2N)$, $O(2N+1)$, and $Sp(2N)$\footnote{Circular Wilson loops and cusp anomalous dimensions for these groups have recently been studied in \cite{Fiol:2014fla}.} have quadratic Casimirs given by \cite{barut} $C_2=2k(k+2\alpha)$, with $\alpha=N-1,N-1/2$, and $N$ in the three respective cases. This will formally lead to the same result as in (\ref{generalization}), with $f=2$, and it also represents a gauge theory prediction, this time for a dual computation of branes on $AdS_5\times \mathbb{RP}^5$, which is the background dual to ${\cal N}=4$ super Yang-Mills with orthogonal or symplectic groups \cite{orientifold}.

There are many natural extensions to our work that are worth being investigated. An obvious one is to find a D3-brane embedding with a non-vanishing geometric cusp $\phi$. Looking at Fig.~\ref{fig-brane_alt}, this would correspond to having the second endpoint of the brane on the AdS$_5$ boundary not at $\psi=\pi$, but at $\psi=\pi-\phi$. Another interesting possibility is to consider the rank $k$ antisymmetric representation. In this case, the dual object would be a D5-brane wrapping an $S^4$ inside of the $S^5$ \cite{DF,Yamaguchi,GP}.


\subsection*{Acknowledgements}
We would like to thank Johannes Henn for helpful comments. DHC and FIS would like to thank  ICTP South American Institute for Fundamental Research for the nice hospitality during the completion of this work. DHC and FIS are supported by CONICET and grants PICT 2012-0417 and PIP 0681. DT would like to thank the Universidad Nacional de La Plata for the nice hospitality during the completion of this work. DT was supported in part by CNPq and by FAPESP grants 2013/02775-0 and 2014/18634-9.


\appendix

\section{Color factors}

In this appendix we collect some formulas for the Casimir operators of $U(N)$ which we have used in the weak coupling analysis of Sec. \ref{perturbative}. For more details the reader is referred to \cite{barut}.

Let $T^a$ (with $a=1, \ldots, N^2$) be the $U(N)$ generators, which can be taken in arbitrary representations  of the group. For any representation ${\cal R}$, they satisfy
\be
{\rm tr}(T^a T^b) = T({\cal R}) \delta^{ab}\,,
\ee
where $T({\cal R})$ is the normalization of the representation. We shall adopt the conventions according to which for the fundamental representation $T(\square) = 1/2$ and define the quadratic Casimir operator through\footnote{The factor $1/2$ was included in this definition in order to have the same expressions for the Casimir coefficients as in \cite{barut}, even though we are using a different normalization for the generators: $(T^a)_\mt{here}=(T^a)_\mt{there}/\sqrt{2}$.}
\be
T^a T^a = \frac{C_2({\cal R})}{2}  {\mathds 1}_{\dim({\cal R})} \,.
\ee

Using the canonical basis, we can replace the $T^a$'s by generators $R^i_j$, which are labeled by two fundamental indices, $i,j=1,\ldots, N$, and satisfy the algebra
\be
[R^i_j,R^k_l] =  \frac{1}{ \sqrt2}\left( \delta_l^i R^k_j- \delta^k_j R_l^i\right)\,.
\label{algebra}
\ee
For example, with our normalization, the generators of the fundamental representation in the canonical basis are given by $(R^i_j)^k_l=\delta^i_l\delta^k_j/\sqrt{2}$. Generic invariant Casimir operators are then defined through
\be
R_{i_1}^{i_2} R_{i_2}^{i_3}\cdots R_{i_p}^{i_1} = \frac{C_p({\cal R})}{2^{p/2}}  {\mathds 1}_{\dim({\cal R})} \,.
\ee
In particular, for totally symmetric or antisymmetric representations of rank $k$ the Casimir coefficients are given by \cite{barut}
\bea
C_p (S_k) = k(N+k-1)^{p-1}\,,
\label{Cp-symm}
\\
C_p (A_k) = k(N-k+1)^{p-1}\,.
\eea
In the main text we have repeatedly used that, for symmetric representations with $k\gg N \gg 1$, the coefficients in (\ref{Cp-symm}) become
\be
C_p(S_k) = k^p\left(1+(p-1)\frac{N}{k}+{\cal O}\left(\frac{1}{k},\frac{N^2}{k^2}\right)\right)\,.
\label{limit-symm}
\ee


\subsubsection*{Perturbative analysis of ladder diagrams}

Ladder diagrams come weighted with color factors which are traces over products of generators. By successive applications of the commutation rules in (\ref{algebra}), all these color factors can be expressed in terms of Casimir coefficients.

The 1-ladder color factor is simply given by the quadratic Casimir
\be
{\rm tr}(T^a T^a) = {\rm tr}(R^i_j R^j_i) = \frac{1}{2}\dim({\cal R}) C_2(\cal R)\,.
\label{base-ind}
\ee

At 2-ladder order, there are two possible color factors
\ba
{\rm tr}(T^a T^a T^b T^b) & = & {\rm tr}(R^i_j R^j_i R^k_l R^l_k) = \frac{1}{4}\dim({\cal R}) C_2({\cal R})^2\,,
\\
{\rm tr}(T^a T^b T^a T^b) & = & {\rm tr}(R^i_j  R^k_l R^j_i R^l_k) = \frac{1}{4}\dim({\cal R}) \left(C_2({\cal R})^2+C_1({\cal R})^2 - N C_2({\cal R})\right)\,,
\ea
the former corresponding to parallel ladders and the latter to crossed ladders, as we have seen in the main text, see Fig. \ref{2loopladders}.

Just to be explicit, let us also consider the 3-ladder order, whose factors are all of the form ${\rm tr}(T^a T^b T^c T^{\sigma(a)}T^{\sigma(b)}T^{\sigma(c)})$.
Whenever $\sigma(a) = c$ or $\sigma(c) = a$, they will be $C_2(\cal R)$ times any of the 2-ladder ones.
The genuinely new factors are the ones with no adjacent generators carrying the same index\footnote{For compactness, we drop the representation label ${\cal R}$ from the Casimirs in these formulas.}
\ba
{\rm tr}(T^a T^b T^c T^a T^b T^c) 
&=& \frac{1}{8} \dim({\cal R}) \left(C_2^3+ 3C_1^2C_2 - 3N C_2^2-2N C_1^2+2N^2 C_2\right)\,,
\\
{\rm tr}(T^a T^b T^c T^a T^c T^b) &=&{\rm tr}(T^a T^b T^c T^b T^a T^c) \,, \cr
&=& \frac{1}{8} \dim({\cal R}) \left(C_2^3+ 2C_1^2C_2 - 2N C_2^2-N C_1^2+N^2 C_2\right)\,.
\ea

At a generic $\ell$-ladder order, the color factors are expressed in terms of higher Casimir coefficients at most up to $C_\ell({\cal R})$, and, more importantly for us, all containing a term that goes like $C_2({\cal R})^{\ell}$. This can be easily seen by induction. Assuming that for any permutation $\sigma$ we have
\be
{\rm tr}(T^{a_1} \cdots T^{a_{\ell}} T^{\sigma(a_1)} \cdots T^{\sigma(a_{\ell})})=
 \dim({\cal R}) \left(\frac{C_2({\cal R})}{2}\right)^{\ell}+\ldots\,,
 \label{orderell}
\ee
we can always use the algebra (\ref{algebra}) and cyclicity of the trace, if convenient, to show that
\be
{\rm tr}(T^{a_1} \cdots T^{a_{\ell+1}} T^{\sigma(a_1)} \cdots T^{\sigma(a_{\ell+1})})=
 \left(\frac{C_2({\cal R})}{2}\right){\rm tr}(T^{a_1} \cdots T^{a_{\ell}} T^{\sigma'(a_1)} \cdots T^{\sigma'(a_{\ell})})+\ldots\,,
 \label{ellipsis}
\ee
by moving a generator $T^{\sigma(a_i)}$ next to a $T^{a_j}$ with $\sigma(a_i)=a_j$. The ellipsis in the formula above contains traces with at most $2\ell+1$ generators, which are produced by the right hand side of (\ref{algebra}) and give rise to terms of the form
\be
\prod_{s=1}^{\ell+1} C_s({\cal R})^{q_s}\,,\qquad
\forall q_s: \quad 0\le q_s\le \ell+1\,,\qquad \sum_{s=1}^{\ell+1} s \,q_s\le 2\ell
\,.
\label{subleading}
\ee
It is easy to see from the constraint above that, in fact, $q_2\le \ell$, so that the ellipsis in (\ref{ellipsis}) does not contain terms that go like $C_2({\cal R})^{\ell+1}$. Together with the base of the induction (\ref{base-ind}), this proves the claim.

An important corollary of this proof is that the color factors for the $\ell$-th ladder diagram for a cusp in the totally symmetric representation with $k\gg N\gg 1$ are all equal to leading order and given by
\ba
{\rm tr}(T^{a_1} \cdots T^{a_\ell} T^{\sigma(a_1)} \cdots T^{\sigma(a_\ell)})
&= & \dim{(S_k)}\left(\frac{k^2}{2}\right)^\ell
\left(1+\ell \frac{N}{k}+{\cal O}\left(\frac{1}{k},\frac{N^2}{k^2}\right)\right)\,.
\ea
Here we have used (\ref{limit-symm}) and (\ref{subleading}) to check that all terms different from $C_2({\cal R})^{\ell}$ in (\ref{orderell}) are subleading in this limit, for they scale at most as $k^{2\ell-2}$. This holds independently of the permutation $\sigma$, thus confirming the statement in (\ref{color-factors}).


\end{document}